\documentclass[%
 aip,
 amsmath,amssymb,
 reprint,%
]{revtex4-1}

\usepackage{graphicx}
\usepackage{dcolumn}
\usepackage{bm}

\usepackage[utf8]{inputenc}
\usepackage[T1]{fontenc}

\makeatletter
\def\@email#1#2{%
 \endgroup
 \patchcmd{\titleblock@produce}
  {\frontmatter@RRAPformat}
  {\frontmatter@RRAPformat{\produce@RRAP{*#1\href{mailto:#2}{#2}}}\frontmatter@RRAPformat}
  {}{}
}%

\preprint{AIP/123-QED}

\makeatother
\begin{document}


\title{Sedimentation of a spheroidal particle in an elastoviscoplastic fluid} 



\author{Alie Abbasi Yazdi}
\affiliation{Dipartimento di Ingegneria Chimica, dei Materiali e della Produzione Industriale, Università degli Studi di Napoli Federico II, P.le Tecchio 80, 80125 Napoli, Italy}

\author{Gaetano D'Avino}
\affiliation{Dipartimento di Ingegneria Chimica, dei Materiali e della Produzione Industriale, Università degli Studi di Napoli Federico II, P.le Tecchio 80, 80125 Napoli, Italy}

\date{\today}

\begin{abstract}
The sedimentation dynamics of a prolate spheroidal particle in an unbounded elastoviscoplastic (EVP) fluid is studied by direct finite element simulations under inertialess flow conditions. The Saramito-Giesekus constitutive equation is employed to model the suspending liquid. The Arbitrary Lagrangian-Eulerian formulation is used to handle the particle motion.

The sedimentation, lift, and angular velocities of spheroids with aspect ratio between 1 and 8 are computed as the initial orientation, Bingham, and Weissenberg numbers are varied. Similarly to the purely viscoelastic case, a spheroid in an EVP fluid rotates up to align its major axis with the applied force. As the Bingham number increases, the settling rate monotonically reduces while the angular velocity first increases and then decreases. The initial orientation has a relevant effect on the particle stoppage because of the different drag experienced by the spheroid as its orientation is varied. The yielded and unyielded regions around the spheroid reveal that, for particle oriented transversely to the force, the yielded envelope shrinks near the tips due to the fast spatial decay of the stresses, and unyielded regions appear along the surface of the particle, similar to the solid caps observed at the front and back of a sphere. Fluid plasticity enhances the negative wake phenomenon that is observed at Weissenberg numbers significantly lower than the purely viscoelastic case. Results of the drag correction coefficient for particles aligned with longest axis along the force are presented.

\end{abstract}


\maketitle 

\section{Introduction}
\label{sec:Introduction}

The dynamics of rigid particles suspended in liquids is strongly affected by fluid rheology. Even in relatively simple flow fields, the non-Newtonian properties of the suspending liquid lead to significant variations of the particle dynamics as compared to Newtonian suspensions \cite{DAvino2015particle}. One of the most widely studied phenomena is the settling of particles in non-Newtonian liquids as it occurs in several industrial and natural processes \cite{Griffiths2000dynamics,Elgaddafi2012settling,Nagasawa2016particle,Frigaard2017bingham}. Experiments, theoretical predictions, and numerical simulations have highlighted the importance of typical non-Newtonian properties as shear-thinning, elasticity, plasticity on the particle settling dynamics, affecting the flow field surrounding the particles and, in turn, the drag coefficient \cite{Beris1985creeping,Jossic2001drag,Mckinley2002transport,Fraggedakis2016yielding}. Despite the obvious relevance, studies on the sedimentation of non-spherical particles in complex liquids are much more limited as compared to spheres. 

Experiments reported that axisymmetric bodies (e.g., cylinders) falling in a viscoelastic fluid at negligible inertia rotate up to reach an orientation with major axis aligned with the force direction \cite{Liu1993sedimentation,Joseph1993orientation}, regardless of fluid rheology, aspect ratio of the particle, and initial orientation. Theoretical calculations and numerical simulations \cite{Leal1975slow,Brunn1977slow,Kim1986motion,Feng1995three,Galdi2002orientation,Dabade2015effects} showed that fluid elasticity generates a non-zero torque on the particle with opposite sign with respect to the one induced by inertia. Studies on spheroidal particles highlighted that the torque is such that both prolate and oblate shapes attain a stable orientation with major axis along the applied force \cite{Dabade2015effects,DAvino2022numerical}. A similar behavior has been found for triaxial ellipsoidal particles where the shape affects the settling dynamics and the steady drag coefficient but not the final orientation \cite{DAvino2022numerical}. Recent numerical simulations have reported the existence of a mastercurve governing the settling dynamics of prolate/oblate spheroids (and a mastersurface for triaxial ellipsoids) after the initial transient due to the viscoelastic stress development, i.e., the settling dynamics only depends on the current orientation regardless of the history to reach
it \cite{DAvino2022numerical}. This behavior is expected whether the characteristic time for the stress development (the fluid relaxation time) is small compared with the characteristic time of the particle rotation, the latter depending on the applied force and the particle shape. 

Recently, a growing interest in the so-called elastoviscoplastic fluids (EVP), i.e., fluids that show both elasticity and a yield stress \cite{Fraggedakis2016yielding1}, is observed. The settling dynamics of a spherical particle in an unbounded EVP liquid has been studied by Fraggedakis et al.\cite{Fraggedakis2016yielding}. The constitutive equation proposed by Saramito \cite{Saramito2007new,Saramito2009new} has been used to model the suspending fluid. The simulation results show typical phenomena observed in viscoelastic fluids such as the negative wake behind the sphere and the loss of the fore–aft symmetry, suggesting that they are due to elasticity and not to thixotropy as previously reported. The paper also proposes a relation for the drag correction coefficient as a function of the Bingham and Weissenberg numbers (see later for the definition) and the limiting values of these parameters for particle stoppage. A quantitative agreement with previous experimental measurements \cite{Holenberg2012particle} is found.

The settling dynamics of a sphere in an EVP liquid in the presence of confinement has been recently addressed. Sarabian et al. \cite{Sarabian2022interface} investigated the sedimentation of a spherical particle placed in the middle of two infinite parallel plates through immersed boundary numerical simulations. As the confinement increases, the velocity of the fluid relaxes faster and the negative wake moves closer to the sphere, resulting in a lower sedimentation rate. An expression of the drag coefficient as a function of the Bingham number and confinement ratio has been also proposed. The sedimentation of a spherical particle in an EVP fluid in proximity of a flat wall has been investigated by Yazdi and D'Avino \cite{Yazdi2023sedimentation} by direct finite element simulations. The Giesekus constitutive equation modified as proposed by Saramito \cite{Saramito2007new} has been used. The results show that the particle migrates orthogonal to the wall and the presence of a yield stress reduces the settling velocity and reverses the migration direction as compared to the purely viscoelastic case. The reversed particle migration phenomenon observed in the elastoviscoplastic fluid is attributed to the different stress distribution around the particle due to the presence of the yielded region.

Few recent studies have investigated the sedimentation of non-spherical particles in viscoplastic (inelastic) fluids. Sobhani et al.\cite{sobhani2019sedimentation} carried out Lattice-Boltzmann simulations to study the sedimentation of a rigid, elliptical (2D) particle in a biviscous, yield stress fluid contained in a finite, closed-ended channel. They reported that the trajectory of the particle is strongly affected by the presence of the yield stress, giving rise to new settling modes. Specifically, a particle initially released at the channel centerline and inclined with respect to the force direction first migrates towards the wall and then moves back towards the centerline, reaching an equilibrium distance from the wall and keeping a non-horizontal orientation while sedimenting. The offset position depends on the Bingham number and the density ratio, and is independent of the initial inclination angle. Da Hui et al.\cite{hui2023sedimentation} employed a graphics processing unit accelerated immersed boundary Lattice-Boltzmann method to study the sedimentation of elliptical particles in a Bingham fluid. The single particle case was considered for validation purposes, confirming the results of Sobhani et al.\cite{sobhani2019sedimentation}. The settling of a pair, a chain, and of a cluster of elliptical particles is addressed, highlighting the appearance of a variety of sedimentation dynamics due to the fluid yield stress. The sedimentation of (3D) prolate and oblate spheroidal particles in a Bingham liquid has been studied by Romanus et al. \cite{Romanus2022fully} by numerical simulations. The particle aspect ratio is varied between 1/4 to 4. Creeping flow conditions are first analyzed, then the study is extended including inertial effects. In the former case, the critical conditions for particle stoppage are evaluated for spheroids oriented with major axis parallel and orthogonal to the force direction. A quantitative match with previous data obtained by the augmented Lagrangian method \cite{Iglesias2020computing} is found. As inertia becomes relevant, the spheroids reach an equilibrium angle depending on the aspect ratio and sediment following a trajectory inclined with respect to the force direction. The analysis of the yielded/unyielded regions reveals that the yielded envelope follows the spheroid extremities during the rotation. A `belt' of plastic material surrounding the particle is found, with dimension depending on the particle shape and orientation. For an inclined prolate spheroid, the plastic belt acts as a lever arm of length dependent on the aspect ratio. For oblate spheroids, two large domes of plastic material form near the flat sides, leading to a stronger rotation resistance. We would like to remark that the just mentioned studies consider (2D or 3D) particles in purely viscoplastic fluids without accounting for elastic effects. To the best of our knowledge, works dealing with the dynamics of non-spherical particles in elastoviscoplastic fluids are not available.

In this work, we study the dynamics of a prolate spheroidal particle settling in an unbounded elastoviscoplastic fluid through direct numerical simulations under inertia-less conditions. The sedimentation, lift, and angular velocities are computed as the relevant fluid, flow, and geometrical parameters are varied. The flow fields around the particle, including the yielded/unyielded fluid regions and the viscoelastic stress distribution, are investigated. Results for the drag correction coefficient as a function of the aspect ratio, Weissenberg and Bingham numbers are provided.

\section{Mathematical model}
\label{sec:Mathematical model}

\subsection{Governing equations}
\label{sec:Governing equations}

\begin{figure*}
\includegraphics[width=0.9\textwidth]{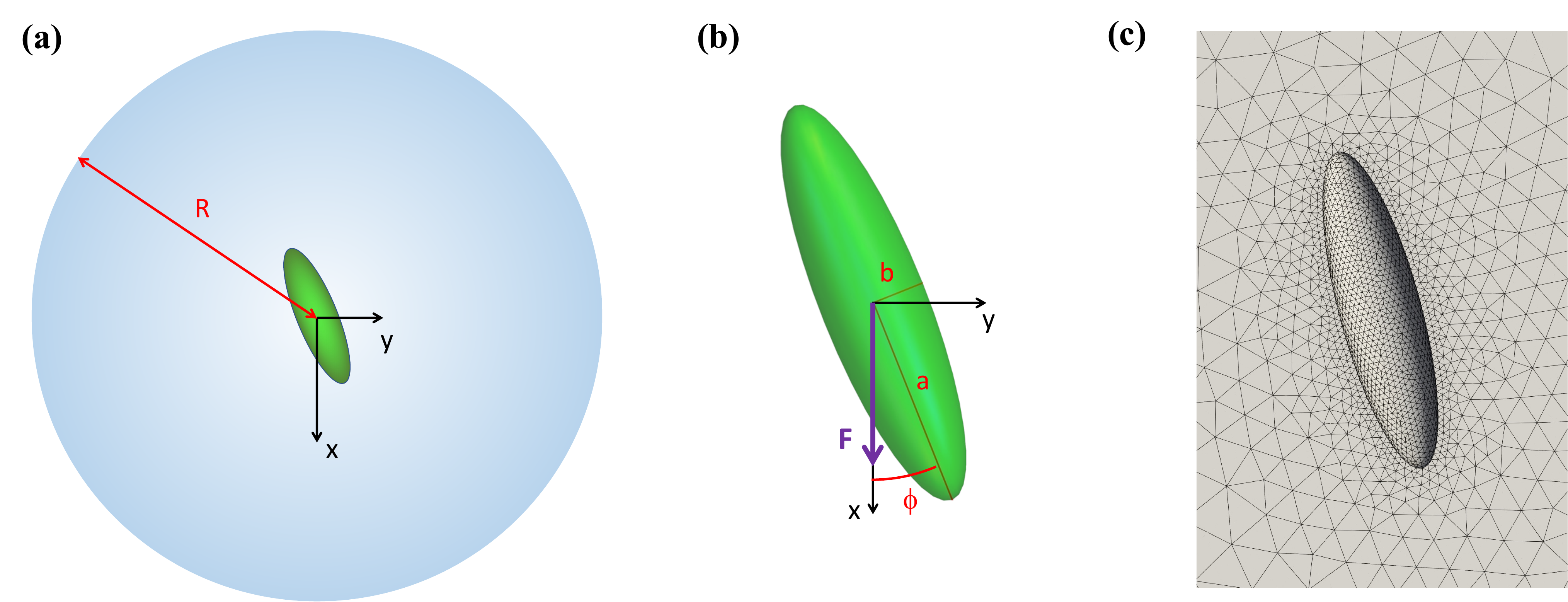}
\caption{(a) Schematic representation of the computational domain: a prolate spheroidal particle is placed at the center of a sphere with radius $R$ with major axis lying on the $xy-$plane of a Cartesian reference frame with origin at the center of the spherical domain. (b) A force $F$ is applied on the particle along the direction of the $x$-axis; the angle between the major axis and the $x-$axis is denoted by $\phi$. (c) Typical mesh around the particle used in the simulations.}
\label{scheme}
\end{figure*}

Figure~\ref{scheme}a displays the computational domain considered in this work. A prolate spheroidal particle with semi-major and semi-minor axes $a$ and $b$ is placed at the center of a spherical domain with radius $R$. As discussed later, the choice of a spherical domain is convenient from a numerical point of view. The domain between the particle and the spherical surface is filled by an EVP fluid. Since our aim is to study the settling dynamics of the particle in an unbounded domain, the radius $R$ is chosen much larger than the characteristic particle size. A Cartesian reference frame is selected with origin at the center of the particle with axes oriented as in figure. A force acting along the $x-$direction is applied on the particle. Without loss of generality, the spheroid major axis is initially placed on the $xy-$plane. Because of the symmetry, the major axis remains on this plane during the sedimentation process and the particle rotates around the $z-$direction. Hence, the angle $\phi$, defined as the angle between the force and the major axis direction (see Fig.~\ref{scheme}b) suffices to describe the particle rotational dynamics.

Assuming inertialess flow conditions, the fluid motion is governed by the continuity and momentum balance equations:
\begin{eqnarray}
	\nabla\cdot \boldsymbol{u}=0
	\label{eqn:continuity},
 \\
	\nabla\cdot\boldsymbol{T}=\boldsymbol{0}
	\label{eqn:momentum}.
\end{eqnarray}
In these equations, $\boldsymbol{u}$ is the fluid velocity and $\boldsymbol{T}$ is the total stress tensor that can be written as:
\begin{eqnarray}
	\boldsymbol{T}=-p\boldsymbol{I}+2\eta_\mathrm{s}\boldsymbol{D}+\boldsymbol{\tau}, \label{eqn:viscoelastic constitutive}
\end{eqnarray}
where $p$ is the pressure, $\boldsymbol{I}$ is the unity tensor, $\eta_\mathrm{s}$ is the viscosity of a Newtonian solvent, $\boldsymbol{D}=(\nabla\boldsymbol{u}+(\nabla\boldsymbol{u})^\mathrm{T})/2$ is the rate-of-deformation tensor, and $\boldsymbol{\tau}$ denotes the extra stress due to the non-Newtonian nature of the fluid and needs to be specified by a constitutive equation. To account for viscoelasticity and viscoplasticity of the suspending liquid, we consider the Giesekus constitutive equation modified as proposed by Saramito \cite{Saramito2007new}:
\begin{equation}
	\lambda\stackrel{\nabla}{\boldsymbol{\tau}}+\max\left(0,\frac{|\boldsymbol{\tau}_\text{d}|-\tau_\text{y}}{|\boldsymbol{\tau}_\text{d}|}\right)\left(\boldsymbol{\tau}+\frac{\alpha\lambda}{\eta_\mathrm{p}}\boldsymbol{\tau}\cdot\boldsymbol{\tau}\right)=2\eta_\mathrm{p}\boldsymbol{D},
	\label{eqn:Saramito_GSK}
\end{equation}
with $\lambda$ the relaxation time, $\eta_\text{p}$ the polymer contribution to the viscosity, $\alpha$ the `mobility' parameter, and $\tau_\text{y}$ the yield stress. The symbol $(^{\nabla})$ denotes the upper-convected time derivative defined as:
\begin{equation}
	\stackrel{\nabla}{\boldsymbol{\tau}}\equiv\frac{\partial\boldsymbol{\tau}}{\partial t}+\boldsymbol{u}\cdot
	\nabla\boldsymbol{\tau}-(\nabla\boldsymbol{u})^{T}\cdot\boldsymbol{\tau}-\boldsymbol{\tau}\cdot
	\nabla\boldsymbol{u}.
 \label{eqn:upper-convected}
\end{equation}
In Eq.~\eqref{eqn:Saramito_GSK}, $|\boldsymbol{\tau}_\text{d}|$ is defined as:
\begin{equation}
	|\boldsymbol{\tau}_\text{d}|=\sqrt{\frac{1}{2}\hat{\boldsymbol{\tau}}:\hat{\boldsymbol{\tau}}},
 \label{eqn:magn_tau}
\end{equation}
and:
\begin{equation}
	\hat{\boldsymbol{\tau}}=\boldsymbol{\tau}-\frac{1}{3}\text{Tr}\boldsymbol{\tau}\,\boldsymbol{I},
	\label{eqn:tau_hat}
\end{equation}
with `Tr' the trace operator. The rheological properties of this constitutive equation in shear and extensional flow have been reported in our previous paper \cite{Yazdi2023sedimentation}.

A force $\boldsymbol{F}$ is applied on the particle along the $x-$direction. Assuming inertialess conditions, the following equations hold at the particle surface $S$:
\begin{eqnarray}
	\boldsymbol{F}=(F,0,0)=\int_S\boldsymbol{T}\cdot \boldsymbol{n}\,dS,\label{eqn:force}
\\
	\int_S(\boldsymbol{x}-\boldsymbol{X})\times(\boldsymbol{T}\cdot\boldsymbol{n})\,dS=\boldsymbol{0}, \label{eqn:torque}
\end{eqnarray}
where $\boldsymbol{X}$ denotes the particle centroid, $\boldsymbol{x}$ is the position vector of a point on the boundary $S$, $\boldsymbol{n}$ is the outwardly directed unit normal vector on $S$ and $dS$ is the local surface area.

Due to the axisymmetric shape of the particle, we can consider one-half of the computational domain where the $xy-$plane passing through the particle center is a symmetry plane. At the external boundaries of the fluid domain (i.e., over the semi-spherical surface in Fig.~\ref{scheme}a) we apply no-slip conditions. The rigid-body motion condition holds on the particle surface:
\begin{equation}
	\boldsymbol{u}=\boldsymbol{U}+\boldsymbol{\omega}\times(\boldsymbol{x}-\boldsymbol{X}), \label{eqn:rigid-body}
\end{equation}
where $\boldsymbol{U}$ and $\boldsymbol{\omega}$ are the translational and angular velocities of the particle. Symmetry conditions are imposed on the symmetry plane. Finally, due to the inertialess assumption, no initial condition for the velocity is needed. The stress-free condition, i.e., $\boldsymbol{\tau}|_{t=0}=\boldsymbol{0}$, is considered as initial condition for the viscoelastic stress.

Once the fluid velocity, pressure and stress fields are calculated along with the particle kinematic quantities, the particle center $\boldsymbol{X}$ and angle $\boldsymbol{\Theta}$ are updated by integrating the following equations:
\begin{eqnarray}
	\frac{d\boldsymbol{X}}{dt}=\boldsymbol{U},~~~\boldsymbol{X}|_{t=0}=\boldsymbol{X}_{\mathrm{0}},
	\label{eqn:kinematic X}
\\
	\frac{d\boldsymbol{\Theta}}{dt}=\boldsymbol{\omega},~~~\boldsymbol{\Theta}|_{t=0}=
	\boldsymbol{\Theta}_{\mathrm{0}},
	\label{eqn:kinematic Theta}
\end{eqnarray}
where $\boldsymbol{X}_{\mathrm{0}}$ and $\boldsymbol{\Theta}_{\mathrm{0}}$ are the initial particle position and orientation. As previously discussed, the symmetry of the particle shape implies that the angular velocity has only one non-zero component, i.e., the rotation around the $z-$axis. Furthermore, the translational velocity has a component along the force direction, that is the sedimentation velocity denoted by $U_\text{s}$, and along the $y-$direction, that is the lift velocity denoted by $U_\text{l}$. Since the domain is unbounded, the initial particle position is irrelevant.

The governing equations can be made dimensionless by selecting proper characteristic quantities. Following our previous work \cite{DAvino2022numerical}, we choose the characteristic length $R_\text{c}$ as the radius of the sphere with the same volume of the spheroid, the characteristic velocity $U_\text{c}$ as the settling velocity of a sphere with radius $R_\text{c}$ in a Newtonian fluid with the same zero-shear viscosity $\eta_0$ of the EVP fluid, and the characteristic stress as $\eta_0 U_\text{c}/R_\text{c}$. The following dimensionless parameters appear:
\begin{gather}
	Wi=\lambda\frac{F}{6\pi \eta_0 R^2} \label{eqn:Wi_number},
 \\
	Bn=\frac{\tau_\text{y}6\pi R^2}{F} \label{eqn:Bn_number},
 \\
	\eta_\mathrm{r}=\frac{\eta_\mathrm{s}}{\eta_\mathrm{0}}, \label{eqn:visc_number}
\end{gather}
that are the Weissenberg number, the Bingham number, and the viscosity ratio, respectively. Along with these three dimensionless parameters, we also have the mobility $\alpha$, the particle aspect ratio $\beta=a/b$, and the initial particle orientation specified by the angle $\phi_0$.

In this work, we fix $\alpha=0.2$ (denoting a shear-thinning fluid) and $\eta_\text{r}=0.1$. The particle dynamics will be studied by varying $Wi$, $Bn$, $\beta$, and $\phi_0$.

\subsection{Numerical method}
\label{sec:Numerical method}

The finite element method is used to solve the governing equations. The DEVSS-G/SUPG formulation \cite{Guenette1995new,Bogaerds2002stability,Brooks1982streamline} and the log-representation for the conformation tensor \cite{Fattal2004constitutive,Hulsen2005flow} have been adopted to improve the convergence at high Weissenberg numbers. The rigid-body motion is imposed through constraints in each node of the particle surface by means of Lagrange multipliers \cite{DAvino2008rotation}. 
As previously discussed, the computational domain used in the simulations is a large sphere with the spheroid at its center. The fluid domain is discretized by tetrahedral elements with a finer resolution close to the particle. Figure~\ref{scheme}c shows a typical mesh around the spheroid. The Arbitrary Lagrangian–Eulerian (ALE) moving mesh method is used to handle the particle motion \cite{Hu2001direct}. The translational velocity of the particle is subtracted from the convective term of the constitutive equation so that the particle center of volume is fixed in space. As previously done \cite{DAvino2014bistability,DAvino2022numerical}, the mesh is rigidly rotated following the particle orientational dynamics, avoiding mesh distortion and remeshing.

The numerical method has been validated in previous publications for spheroidal particles in viscoelastic fluids \cite{DAvino2014bistability,DAvino2022numerical} and for spherical particles in EVP fluids \cite{Yazdi2023sedimentation}. In this work, we have verified mesh and time convergence for all the investigated parameters. The required mesh resolution and time step size to get convergent results depend on $Wi$, $Bn$, and the particle aspect ratio. We found that about $50000$ tetrahedral elements assure spatial-independent solutions for the most critical cases. The (dimensionless) typical time step size is 0.005 that must be reduced to 0.001 for the highest values of $Wi$ and $Bn$. Finally, the value of the radius $R$ of the spherical domain is a delicate choice. Indeed, as pointed out by Moschopoulos et al. \cite{Moschopoulos2021concept}, because of the adopted constitutive equation, a variation of the velocity field in the unyielded area travels through the domain up to reach the external boundary, disturbing the numerical solution. Hence, we selected a very large domain ($R=100R_\text{c}$) so that a stationary condition is reached before the numerical solution is affected by this issue.

\section{Results}
\label{sec:Results}

\begin{figure*}
\includegraphics[width=1.0\textwidth]{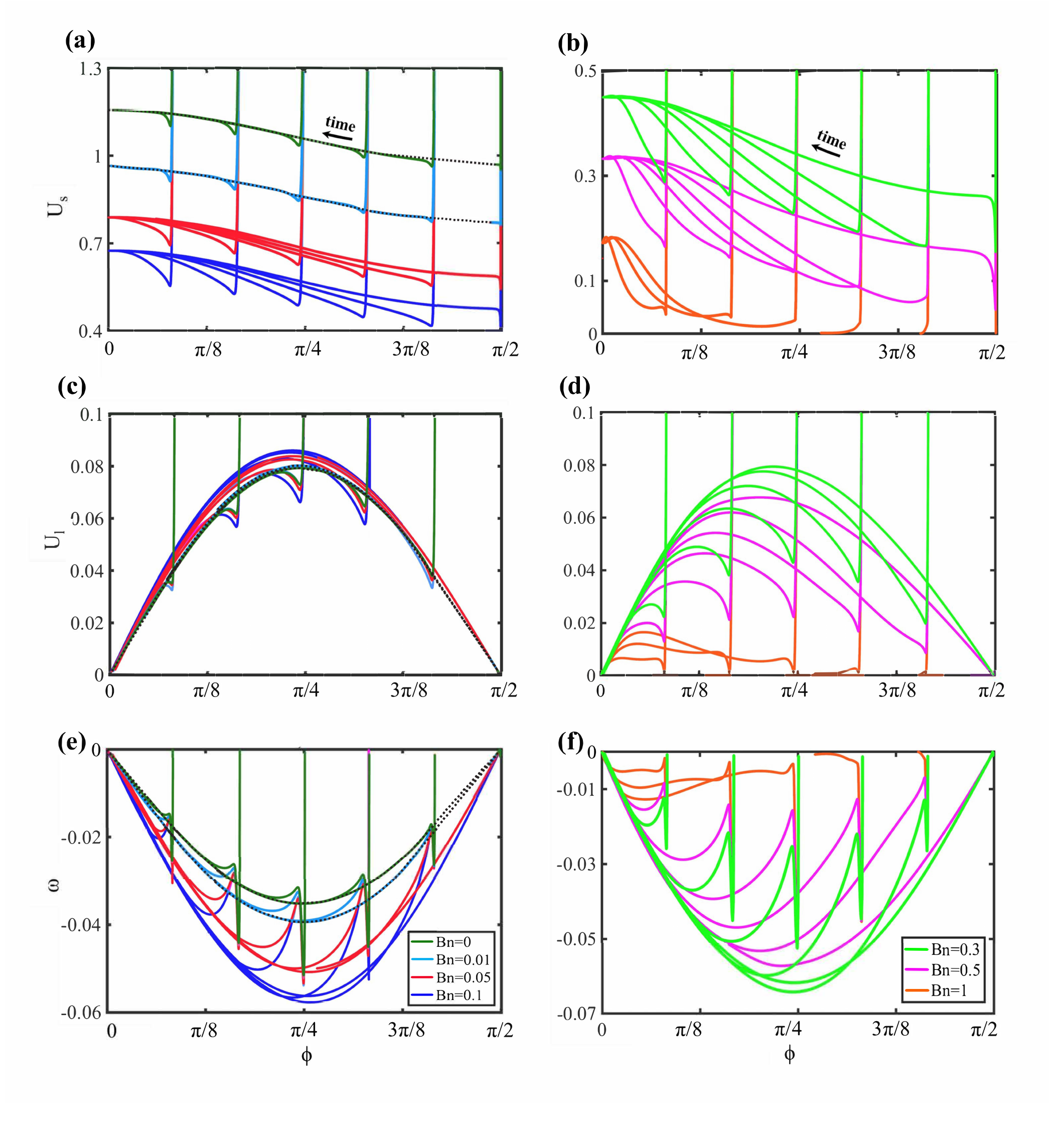}
\caption {Sedimentation velocities (a-b), lift velocities (c-d), and angular velocities (e-f) of a prolate spheroid with an aspect ratio $\beta = 2$ settling in an elastoviscoplastic fluid at $Wi = 1$ as a function of the orientation angle and for different values of the Bingham number. The curves correspond to different initial orientations.}
\label{beta2}
\end{figure*}

The dynamics of a spheroidal particle in an EVP fluid is presented in terms of the translational and rotational velocities as the aspect ratio, initial orientation, Weissenberg and Bingham numbers are varied. Simulation results for the purely viscoelastic case are also included to highlight the effect of fluid plasticity on the particle motion. The diagrams presented in Fig.~\ref{beta2} show the sedimentation velocity $U_\text{s}$ (the velocity component along the force direction), the lift velocity $U_\text{l}$ (the velocity component orthogonal to the force direction), and the angular velocity $\omega$ around the $z$-axis as a function of orientation angle $\phi$ of a prolate spheroid with aspect ratio $\beta=2$ and $\text{Wi}=1$ for different values of the Bingham number. Each curve represents a different initial orientation, denoted by $\phi_0$, changing from $0$ to $\pi/2$ with $\pi/12$ increment. For a better visualization of the results, we split the curves in two plots: the left column of Fig.~\ref{beta2} reports the data up to $Bn=0.1$ and the right one from $Bn=0.3$ up to $Bn=1$. Each curve shows the evolution of the particle kinematic quantities as a function of the particle orientation, i.e., time runs along the curves. The first point of each curve corresponds to a spheroid in a Newtonian fluid under creeping flow conditions due to the stress-free initial condition \cite{DAvino2022numerical}. Because of the small solvent viscosity contribution used in this work, the sedimentation and lift velocities are initially very large (and beyond the selected range for the $y-$axes of the diagrams) whereas the angular velocity is zero. Similarly to the purely viscoelastic case, the kinematic quantities rapidly change due to the build-up of the viscoelastic stresses, as denoted by the almost vertical trends of the curves. Then, depending on the Bingham number, different behaviors can be observed. 

\begin{figure*}
\includegraphics[width=0.9\textwidth]{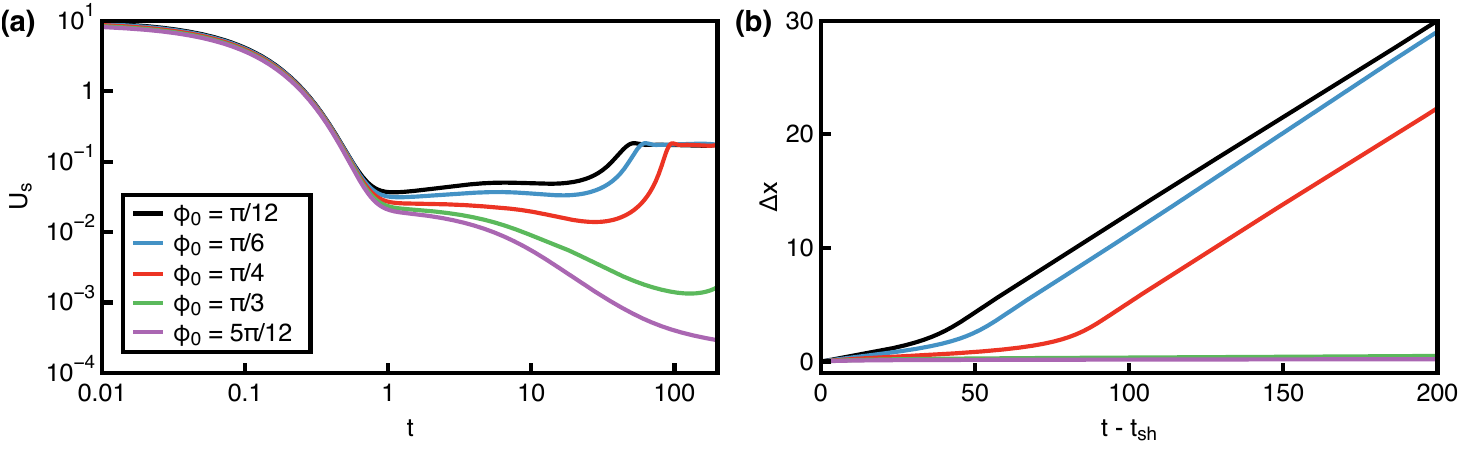}
\caption{(a) Sedimentation velocity and (b) distance travelled by the spheroid along the force direction as a function of time for a particle with aspect ratio $\beta=2$, $Bn=1$, $Wi=1$ for different initial orientation angles. In panel (b), $t_\text{sh}=2$ is the time needed to almost completely extinguish the initial stress build-up.} 
\label{transient_beta2}
\end{figure*}

\begin{figure*}
\includegraphics[width=1.0\textwidth]{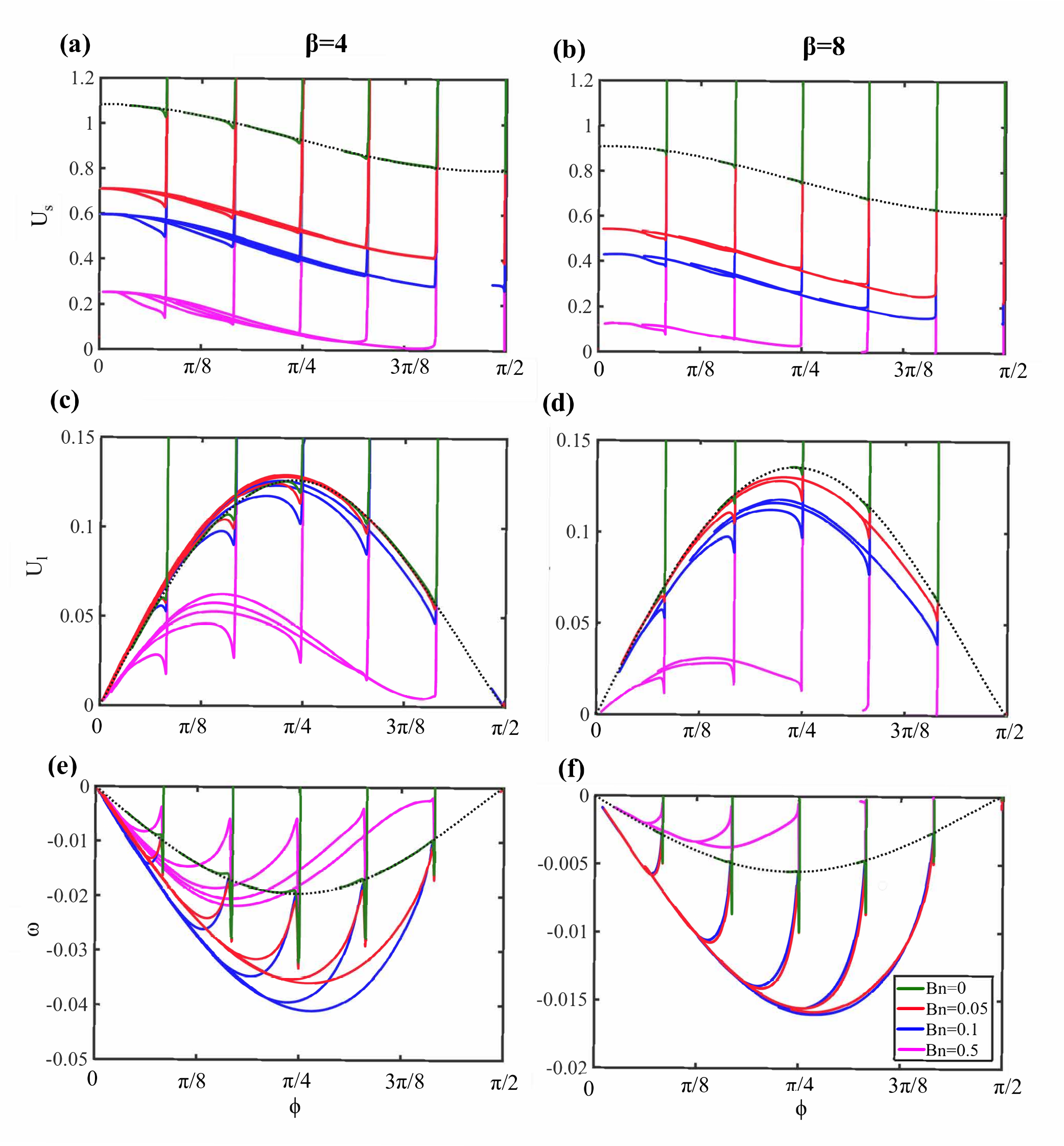}
\caption{Sedimentation velocities (top), lift velocities (middle), and angular velocities (bottom) of a prolate spheroid settling in an elastoviscoplastic fluid at $Wi=1$, $\beta=4$ (left) and $\beta = 8$ (right) as a function of the orientation angle. The curves correspond to different initial orientations.}
\label{beta4_8}
\end{figure*}

First of all, for all the cases, the angular velocity is negative denoting a clockwise rotation leading the spheroid to align with major axis along the force direction. Furthermore, the sedimentation velocity increases from $\phi=\pi/2$ to $\phi=0$, i.e., the fastest sedimentation rate corresponds to the spheroid aligned with the longest axis along the applied force. For a purely viscoelastic case ($Bn=0$), once the elastic stresses fully develop, the curves overlap indicating the presence of a mastercurve denoted by the dotted lines in Fig.~\ref{beta2} \cite{DAvino2022numerical}. The same behavior is observed at $Bn=0.01$. In this case, the sedimentation velocity is lower than the viscoelastic case due to the effect of fluid plasticity. The lift velocity is only barely affected, whereas the angular velocity increases in magnitude. At $Bn=0.05$ and $Bn=0.1$, the settling velocity further reduces and the trends for different initial orientations visibly do not overlap anymore. While the lift velocity does not significantly change, the angular velocity further increases in magnitude. Hence, from $Bn=0$ to $Bn=0.1$, the spheroid settles slower but rotates faster, reaching the stable orientation in a shorter time. Notice also that all the angular velocity curves in Fig.~\ref{beta2}e show an undershoot followed by a overshoot in a very short time (in about one dimensionless relaxation time). After the overshoot, the angular velocity of the purely viscoelastic case remains quantitatively similar to the overshoot itself, whereas it becomes larger and larger as $Bn$ increases. A larger $\omega$ implies that, during the transient of the stress build-up, the spheroid changes its orientation faster at high $Bn$ and the curves overlap at later times or not overlap at all. The faster rotation rate at high $Bn$-values is likely induced by the existence of the yielded region surrounding the particle that acts as a confinement. As discussed later, such a region is not symmetric and changes shape with the particle orientation.

The behavior just discussed is even more pronounced at higher Bingham numbers as shown in the right column panels of Fig.~\ref{beta2}. At $Bn=0.3$, the magnitude of the angular velocity further increases and, indeed, the yellow curves overlap only when the spheroid has almost reached the equilibrium orientation. A similar behavior is observed at $Bn=0.5$. However, in this case, the angular velocity is starting decreasing as the yielded region is approaching the particle (see later). Despite the lower angular velocity, the curves at different initial orientations follow different trends and overlap only when the spheroid almost aligned along the force. This behavior is due to the longer and longer transient dynamics observed at high Bingham numbers.

At $Bn=1$, all the kinematic quantities significantly reduce since the particle is going towards the stoppage condition. It is interesting to note that the curves corresponding to initial orientation $\phi_0=\pi/4$ or lower show a similar trends to those at smaller $Bn$-values, i.e., the particle (slowly) rotates to the vertical equilibrium position with an increasing sedimentation rate; on the contrary, for $\phi_0=\pi/3$ or higher, all the kinematic quantities (in magnitude) are close to zero so the settling dynamics is extremely slow or even the particle is expected to be blocked at some orientation. All the red curves correspond to the same final simulation time (200 dimensionless times). Due to the very low angular velocity, the curve at $\phi_0=\pi/3$ ends before the others since the particle changes its orientation $\phi$ very slowly. A more detailed investigation on the dynamics at $Bn=1$ is reported in Fig.~\ref{transient_beta2}a where the sedimentation velocity $U_\text{s}$ is plotted as a function of time for different initial orientation angles. After the initial transient due to the stress build-up, the curves for $\phi_0 \le \pi/4$ reach the same plateau corresponding to the vertical alignment. At $\phi_0=\pi/3$ (green curve), the settling rate achieves a minimum at around $t=100$ and then starts increasing. At longer times, we expect that particle rotates to attain the stable vertical orientation and the sedimentation velocity curve reaches the plateau observed at lower $\phi_0$. However, since the minimum of $U_\text{s}$ is two orders of magnitude lower than the steady-state value, the time needed to achieve the stable orientation is very long. This is confirmed in Fig.~\ref{transient_beta2}b where the dimensionless distance travelled by the spheroid along the force direction is reported as a function of time. To neglect the effect of the initial viscoelastic stress transient, the time is shifted by $t_\text{sh}$ (set equal to 2) which is the time needed to almost completely extinguish the stress build-up. The green curve at $\phi_0=\pi/3$ remains nearly horizontal within the investigated time window denoting no appreciable motion of the particle, in contrast with the curves at lower $\phi_0$ where the travelled distance is up to 30 times the characteristic particle size. At $\phi_0=5\pi/12$ (purple curves in Fig.~\ref{transient_beta2}), the settling rate is one order of magnitude smaller at long times, further slowing down the sedimentation dynamics. We cannot assess whether the settling velocity rises towards the steady-state value. If this is the case, however, we expect that the time to achieve an appreciable settling rate is so long that the particle can be considered blocked for any practical purpose.

In summary, at $Bn=1$, a critical initial orientation angle $\phi_\text{0,cr}$ can be identified such that particles released at $\phi_0 < \phi_\text{0,cr}$ (slowly) settle and rotate towards the stable vertical orientation. Such particles coexist with spheroids released at $\phi_0 > \phi_\text{0,cr}$ that experience a larger drag and are blocked in the fluid (or settle extremely slow). For these particles, the instantaneous Bingham number, i.e., the Bingham number defined with the current sedimentation velocity, is lower than the critical value for particle stoppage and the spheroid is not able to settle and rotate. Particles with major axis closer to the force direction settle sufficiently fast to overcome the critical Bingham number and follow the expected sedimentation dynamics. By further increasing $Bn$, the critical initial orientation reduces and more and more particles oriented with major axis near the force are blocked. Above a critical Bingham number $Bn_\text{cr}$, all the spheroids stop translating and rotating regardless of the initial orientation. Such a critical value is the one corresponding to the blockage of the spheroid with longest axis along the force, and will be discussed later. 

\begin{figure*}
\includegraphics[width=1.0\textwidth]{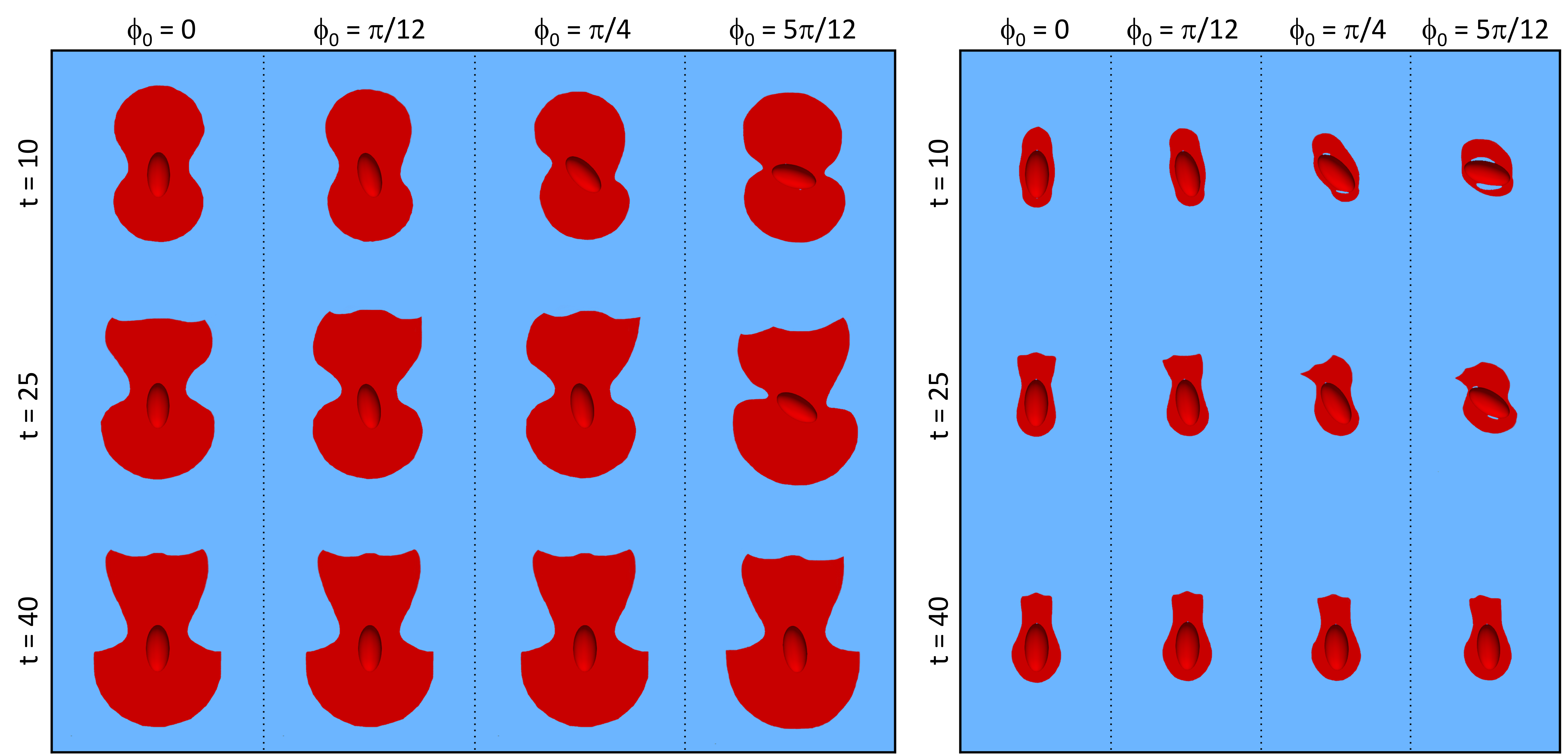}
\caption{Snapshots of the yielded (red) and unyielded (cyan) regions on the symmetry plane around a spheroid with aspect ratio $\beta=2$ and $Wi=1$. Four values of the initial orientation angle and three (dimensionless) time instants are considered. The left and right images refer to $Bn=0.1$ and $Bn=0.5$, respectively.}
\label{yielded_beta2}
\end{figure*}

\begin{figure*}
\includegraphics[width=1.0\textwidth]{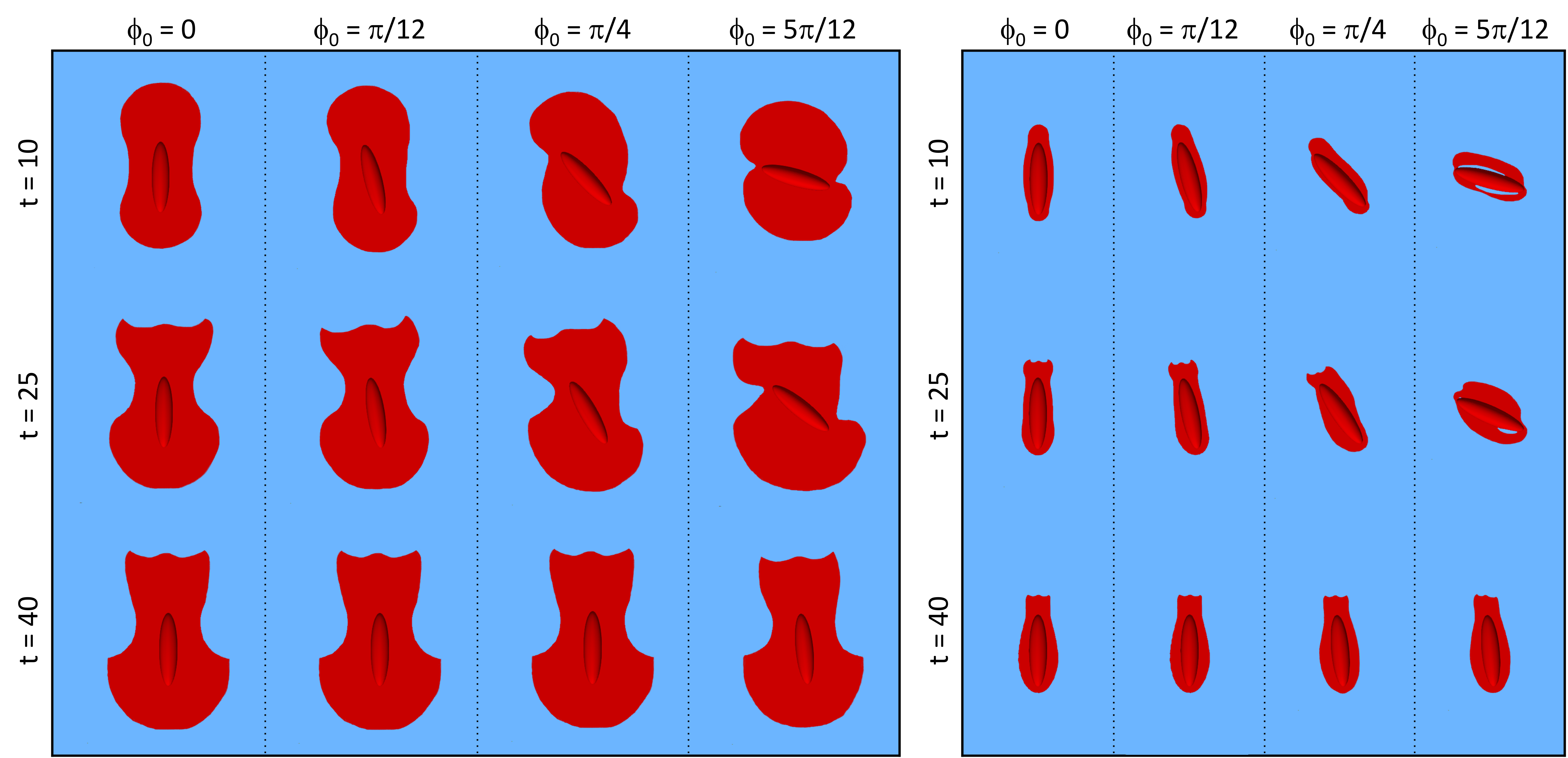}
\caption{Snapshots of the yielded (red) and unyielded (cyan) regions on the symmetry plane around a spheroid with aspect ratio $\beta=4$ and $Wi=1$. Four values of the initial orientation angle and three (dimensionless) time instants are considered. The left and right images refer to $Bn=0.1$ and $Bn=0.5$, respectively.}
\label{yielded_beta4}
\end{figure*}

\begin{figure*}
\includegraphics[width=0.9\textwidth]{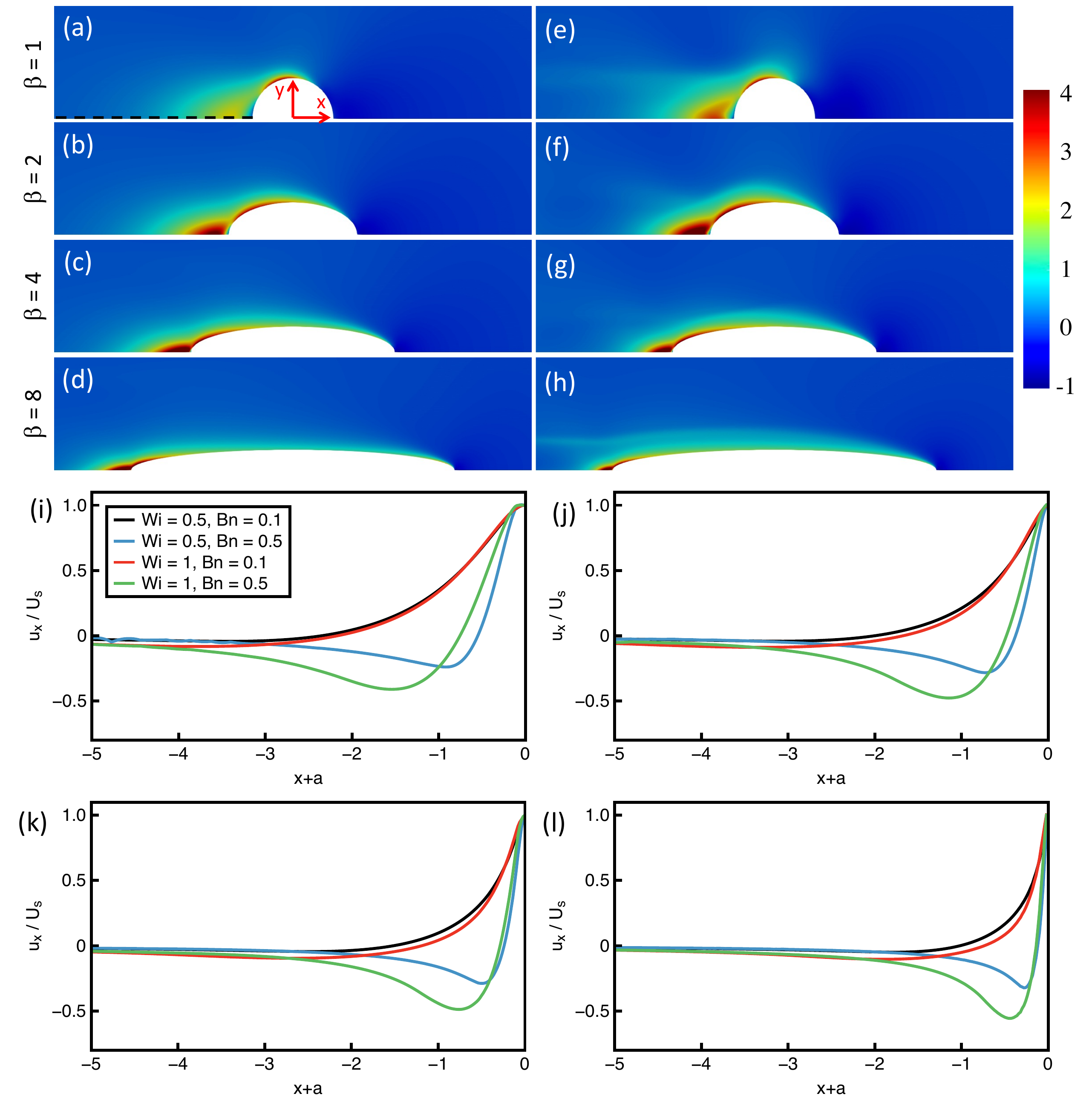} 
\caption{ (a)-(h) Distribution of $\tau_\text{xx}$ around a spheroid with the major axis aligned with the force direction at two Bingham numbers $Bn=0.1$ (left column) and $Bn=0.5$ (right column), and four aspect ratios. The Weissenberg number is $Wi=1$. (i)-(l) $x$-component of the fluid velocity normalized by the steady-state settling velocity along the axis of symmetry behind the particle indicated as the dashed line in panel (a). In panels (i)-(l), the axial coordinate $x$ is translated by the semi-major axis, thus the point at $x=0$ coincides with the particle tip.}
\label{negative_wake}
\end{figure*}

The effect of aspect ratio on the settling dynamics is shown in Fig.~\ref{beta4_8} where the sedimentation, lift and angular velocities of spheroids with aspect ratio $\beta=4$ and $\beta=8$ are reported for $Wi=1$. As the aspect ratio increases, the settling velocity reduces for a fixed Bingham number. Whereas for $\beta=4$ the existence of a mastercurve is visible only at relatively low $Bn$-values (like the case at $\beta=2$), the curves for $\beta=8$ at different initial orientations fairly overlap up to $Bn=0.5$. Indeed, at high aspect ratios, the particle rotates slower (compare the angular velocity values between Fig.~\ref{beta4_8}f, Fig.~\ref{beta4_8}e and Figs.~\ref{beta2}e-\ref{beta2}f) and the stress build-up ends while the spheroid did not significantly change its orientation. Because of the enhanced drag, a spheroid with higher aspect ratio stop translating and rotating at a lower Bingham number. This is clearly visible for $\beta=8$ where, at $Bn=0.5$ (purple curves), spheroids that (slowly) rotate towards the stable orientation coexist with blocked particles. 

The settling dynamics of a spheroid with different aspect ratio is also studied at $Wi=0.5$. The results (not reported) show that, as the Weissenberg number decreases, both the translational and rotational velocities reduce, leading to a slower overall settling dynamics. Because of the smaller angular velocity and the shorter characteristic time for the stress build-up, mastercurves are observed for all the investigated cases. The trends of the kinematic quantities are qualitatively similar to those at $Wi=1$ and only quantitative differences are observed. Specifically, the slower settling velocity leads to a smaller critical Bingham number for particle stopping.

The shape and extension of the yielded/unyielded zones surrounding a particle in elastoviscoplastic suspensions, i.e., the regions in the fluid domain where $|\tau_d|$ is higher (indicating fluid behavior) or lower (indicating solid behavior) than the yield stress, is an intriguing characteristic of yield stress materials. Figure~\ref{yielded_beta2} displays the yielded (red) and unyielded (cyan) regions around a spheroidal particle on the $xy$-plane (the symmetry plane). The Weissenberg number is fixed at $Wi=1$, and the aspect ratio is $\beta=2$. Panels (a) and (b) refer to Bingham numbers of $Bn=0.1$ and $Bn=0.5$, respectively. Three time instants and four initial particle orientations are considered. The same plots for a spheroid with aspect ratio $\beta=4$ are shown in Fig.~\ref{yielded_beta4}.

As expected, from $Bn=0.1$ to $Bn=0.5$ the yielded region reduces. For particles aligned with the force direction, the shape of the yielded region is qualitatively similar to spheres (see, e.g., Fraggedakis et al. \cite{Fraggedakis2016yielding}) with the loss of the fore-aft symmetry due to the fluid elasticity. As pointed out by Romanus et al. \cite{Romanus2022fully}, the shape of the yielded region changes during the settling dynamics, following the particle orientation. Specifically, for particles oriented transversely to the flow direction, the yielded region `shrinks' towards the tips of the spheroid, as visible in the snapshots at $\phi_0=5\pi/12$ for $Bn=0.1$. This phenomenon is likely related to the fact that, because of the elongated shape, the fluid stress is enhanced in the region very close to the tip but decays faster far from the particle, as observed for the purely viscoelastic case \cite{DAvino2022numerical}. Hence, as the particle is more elongated, the fluid stress becomes lower than the yield stress at a shorter distance from the tip. This behavior is confirmed in Fig.~\ref{yielded_beta4} where, for $\phi_0=5\pi/12$, the unyielded region is almost in contact with the particle tips until the spheroid major axis is nearly perpendicular to the force direction. At $Bn=0.5$, the effect of the particle orientation is even more relevant. Indeed, for spheroids nearly oriented along the force direction, the yielded region surrounds the particle similarly to the case at $Bn=0.1$. On the contrary, the yielded fluid zone around particles with major axis transversal to the force contains unyielded regions, as visible in the rightmost snapshots in Figs.~\ref{yielded_beta2} and \ref{yielded_beta4}. These embedded unyielded regions are rather thin and extend along the flatter surface of the spheroid. During the settling dynamics, the particle rotates to reach the vertical stable orientation and these zones reduce up to disappear at some orientation. These regions are similar to the solid caps at the front and back of a spherical particle where the stress is very low due to the stagnation point \cite{Beris1985creeping}. For an elongated particle oriented transversal to the flow, the low stress region is more extended as compared to spheres and the solid regions are wider. We do not observe embedded unyielded regions for vertically aligned particles as the stresses around the particle tip are large.

The results just discussed show that, if the spheroid is not blocked at some orientation, the angular velocity is such that the particle, while settling, rotates to align the major axis along the force regardless of the aspect ratio, Weissenberg and Bingham numbers. In the rest of the paper, we provide additional results assuming that the particle has reached the equilibrium orientation. In this regards, we adopt an axisymmetric version of our code that allows to explore a wide range of parameters at reduced computational cost. The results in terms of particle velocities, yielded/unyielded region, fluid velocity and stress fields, are preliminary compared with those obtained by the 3D version of the code under same conditions showing an excellent quantitative agreement.

Figures~\ref{negative_wake}a-\ref{negative_wake}h display the color map of the axial component of the viscoelastic stress $\tau_\text{xx}$ around the spheroid (the particle is settling from the left to the right) for different aspect ratio at $Wi=1$. The left and right panels refer to $Bn=0.1$ and $Bn=0.5$, respectively. Similarly to the viscoelastic case, as the aspect ratio increases, the axial stress reduces along the particle side, while it increases in the wake due to the high curvature of the shape, denoting a stronger extensional flow behind the particle. The extension of the high stress region, however, decreases as the particle is more elongated. Interestingly, the effect of the Bingham number depends on the aspect ratio: at $\beta=1$ and $\beta=2$, higher Bingham numbers increase $\tau_\text{xx}$ in the particle wake, whereas, at $\beta=4$ and $\beta=8$, the opposite is found. 
 
\begin{figure}
\includegraphics[width=0.45\textwidth]{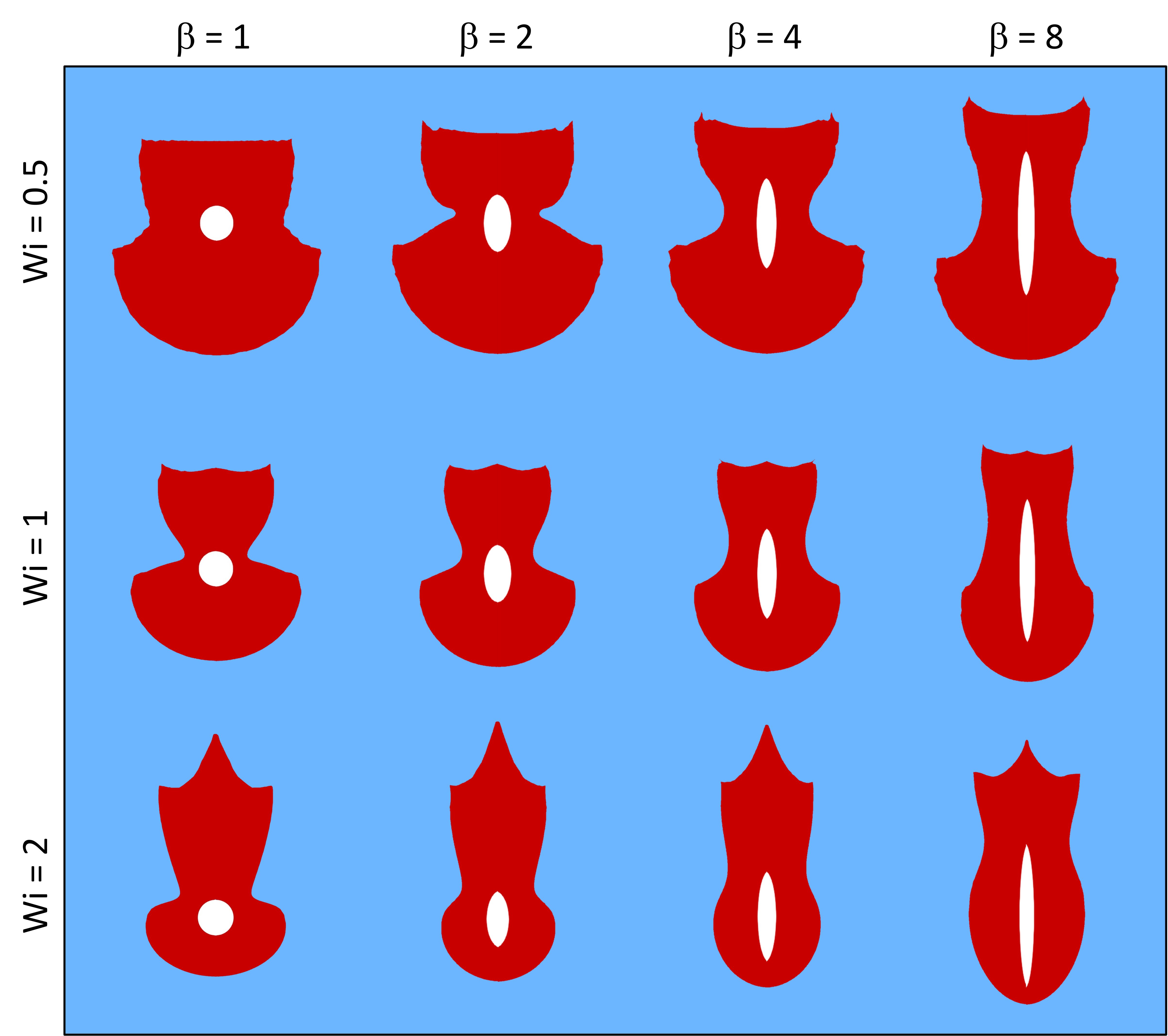}
\caption{Snapshots of the yielded (red) and unyielded (cyan) regions on the symmetry plane around a spheroid aligned with the force direction for different aspect ratios and Weissenberg numbers. The Bingham number is $Bn=0.1$.}
\label{yielded_axis}
\end{figure}

\begin{figure*}
\includegraphics[width=0.8\textwidth]{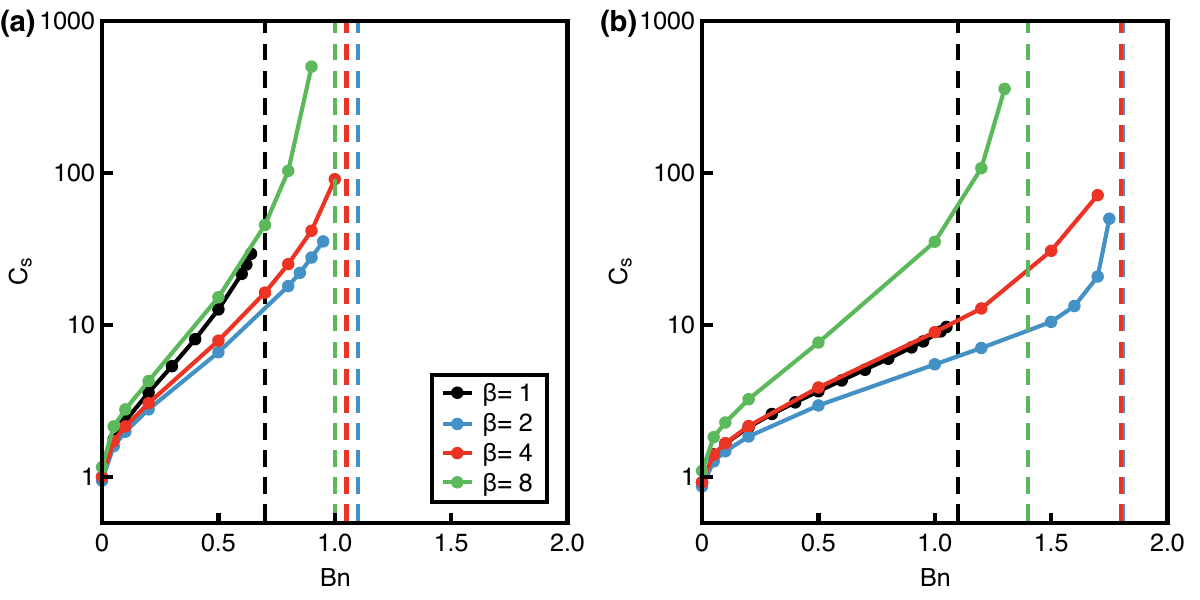}
\caption{Drag correction coefficient as a function of the Bingham number for different values of the aspect ratio. The Weissenberg number is $Wi=0.5$ in panel (a) and $Wi=1$ in panel (b).}
\label{critical_Bn}
\end{figure*}

As reported for viscoelastic fluids \cite{Dou2004criteria,DAvino2022numerical}, higher values of $\tau_\text{xx}$ and its axial gradient lead to a more rapid variation of the fluid axial velocity, possibly leading to the negative wake phenomenon \cite{Harlen2002negative}. Figures~\ref{negative_wake}i-\ref{negative_wake}l show the fluid velocity $u_x$ normalized by the steady-state particle settling velocity along the symmetry axis (the $x$-coordinate of the graphs is translated by the particle semi-major axis $a$ so that the rightmost point identifies the particle tip). Four combinations of $Wi$ and $Bn$ are considered. We recall that the fluid behind a sphere/spheroid in a purely viscoelastic case does not show a negative velocity up to $Wi=2$ \cite{DAvino2022numerical}. On the contrary, the negative wake phenomenon is observed for all the examined cases. Even at $Wi=0.5$ and $Bn=0.1$ (black curves), the fluid velocity is slightly negative in a wide region behind the particle, regardless of the aspect ratio. The negative wake phenomenon is enhanced at $Wi=1$ although the effect is relatively small. A much stronger negative velocity is observed as $Bn$ is increased to 0.5 (blue and green curves). In this case, the minimum fluid velocity is about one-half of the settling speed. As the aspect ratio increases, such a minimum is (in magnitude) larger and moves towards the particle tip.

The faster decay of the stresses and the consequent shrinkage of the high stress region in the wake of the particle discussed above affect the extension of the yielded region behind the particle. Such effect, partially shown in Figs.~\ref{yielded_beta2} and \ref{yielded_beta4}, is clearly visible in Fig.~\ref{yielded_axis} where the yielded/unyielded regions are compared for different Weissenberg numbers and aspect ratios. The Bingham number is fixed to $Bn=0.1$. As $\beta$ increases, the yielded interface behind the spheroid approaches the particle surface. A similar behavior occurs at the front of the particle. The shape of the yielded region shows the same features as the aspect ratio is varied, with the lateral concavity that is smoother and moves backwards as the particle is more elongated. Figure~\ref{yielded_axis} displays the effect of the Weissenberg number as well. As $Wi$ increases, the yielded envelope is closer to the front of the particle surface and farther from its back. This is a consequence of the different stress region around the particle as $Wi$ is varied \cite{DAvino2022numerical}. Specifically, at higher Weissenberg values, the stresses behind the particle increase in magnitude and decay slower. Hence, a wider fluid region experiences a stress higher than the yield stress. The opposite happens ahead the particle where the stress decreases in magnitude and decays faster as $Wi$ increases, thus reducing the extension of the yielded region. Increasing the Weissenberg number also changes the shape of the yielded envelope, leading to a narrower fluid region alongside the spheroid and the appearance of a central `horn' on the axis of symmetry behind the particle.

We finally investigate the drag correction coefficient defined as $C_\text{s}=F/(6\pi\eta_0 R_\text{c}U_\text{s})$ representing the deviation from the Stokes drag (valid for an isolated sphere in a Newtonian fluid under creeping flow conditions). Figure~\ref{critical_Bn} reports $C_\text{s}$ as a function of the Bingham number for four values of the aspect ratio and for $Wi=0.5$ and $Wi=1$. The point at $Bn=0$ denotes the purely viscoelastic case which is slightly lower than 1 as elasticity reduces the drag compared to a Newtonian fluid \cite{Su2022data,DAvino2022numerical}. The points of the curves correspond to the values of the drag correction coefficient once a steady-state is reached. In this regards, we found that, as the Bingham number increases, the time needed to reach a stationary condition increases as well. The vertical dashed lines denote the lowest values of the Bingham number such that, after 5000 dimensionless times, the sedimentation velocity did not reach a steady-state but decreases in time, similarly to the purple curve in Fig.~\ref{transient_beta2}, and reaches values lower than $10^{-5}$. As previously commented, we cannot assess whether the trend will increase at longer times. In any event, the sedimentation phenomenon is extremely slow so that the particle can be considered stopped for practical applications. Our numerical method is not able to directly compute the exact value of the critical Bingham number corresponding to the stoppage condition and special techniques are required \cite{Iglesias2020computing}. However, we expect that the critical Bingham number does not differ too much from the values denoted by the dashed lines. For $Bn-$values between the rightmost point of the curves and the dashed lines, we found that a steady-state is not reached in 5000 dimensionless times, although the sedimentation rate attained or is attaining a plateau and, then, is likely to rise so the corresponding $Bn$ is lower than the critical value. Regardless of the aspect ratio, the drag increases for increasing values of the Bingham number, i.e., the particle sediments slower.  The drag correction coefficient and the values of the critical Bingham number are a non-monotonic function of the aspect ratio, decreasing from the sphere to the spheroid with $\beta=2$ and then increasing for more elongated particles. This result agrees with the Newtonian and viscoelastic cases where the minimum drag coefficient is observed for a prolate spheroid with aspect ratio of about 2. Hence, as previously discussed, a more elongated particle gets stopped at a lower Bingham number. Finally, from $Wi=0.5$ to $Wi=1$, the drag correction coefficient reduces for a fixed value of $Bn$ and the critical Bingham number is larger due to the aforementioned drag reduction induced by fluid elasticity. 

\section{Conclusions}
\label{sec:Conclusions}

The settling dynamics of a prolate spheroidal particle in an unbounded elastoviscoplastic fluid is studied by direct numerical simulations under inertialess flow conditions. The Saramito-Giesekus constitutive equation is employed to model the suspending liquid. The equations governing the fluid and particle dynamics are solved through the finite element method. The Arbitrary Lagrangian-Eulerian formulation is used to handle the particle motion.

Similarly to the purely viscoelastic case, a spheroid in an EVP fluid rotates up to attain a stable orientation with major axis along the force direction, corresponding to the fastest sedimentation. As the Bingham number increases, the settling rate reduces. On the contrary, the angular velocity first increases (in magnitude) and then decreases. This non-monotonic behavior might be attributed to the yielded region surrounding the particle that acts as a confinement, speeding up the rotational dynamics. However, as the Bingham number further increases, the yielded region approaches the particle leading to the stoppage condition, thus slowing down the rotation rate.

The initial orientation has a relevant qualitative effect on the settling dynamics when the Bingham number is close to the critical one, leading to the coexistence of particles initially oriented with the longest axis close to the direction perpendicular to the force that experience a large drag and are not able to translate/rotate, and particles with major axis more aligned with the force which (slowly) rotate towards the stable orientation. As the aspect ratio increases, the drag experienced by the particle increases too, thus the spheroid stops translating and rotating at a lower Bingham number. Also, a more elongated particle rotates slower and the stress build-up ends while the spheroid did not change too much its orientation, leading to the appearance of a mastercurve for the kinematic quantities as a function of the current particle orientation, like for the purely viscoelastic case. At higher Weissenberg numbers, the drag reduces speeding up the particle sedimentation and moving the critical Bingham number at higher values.

The shape of the yielded region depends on the Bingham number and the aspect ratio, and changes as the particle rotates. For particles oriented transversely to the force, the yielded region shrinks near the tips of the spheroid due to the fast spatial decay of the stresses in that regions. As the Bingham number increases, unyielded regions appear along the surface of the particles, similar to the solid caps observed at the front and back of a sphere. While rotating towards the stable orientation, these unyielded regions disappear as the stresses behind the spheroid are larger due to the particle tip. Such larger stresses are responsible for the negative wake phenomenon that, in an EVP fluid, is observed at Weissenberg number values significantly lower than the purely viscoelastic case.

The results presented in this work refer to an unbounded system. The presence of a confining wall strongly affects the particle dynamics as reported for spheres in an EVP material \cite{Chaparian2020particle,Sarabian2022interface,Yazdi2023sedimentation}. Future work will be dedicated to investigate the dynamics of spheroidal particles in an EVP liquid including the effect of confinement, with a special interest on pressure-driven channel flows and sedimentation near a wall container. 

\begin{acknowledgments}
This work is carried out in the context of the project YIELDGAP (https://yieldgap-itn.eu) that has received funding from the European Union’s Horizon 2020 research and innovation program under the Marie Skłodowska-Curie grant agreement No 955605.
\end{acknowledgments}


\bibliography{bibliography.bib}

\end{document}